\documentclass[a4paper,11pt]{article}
\usepackage{jheppub} 
\usepackage{lineno}
\usepackage{main-defs}

\title{\boldmath Semi-supervised permutation invariant particle-level anomaly detection}

\author[a]{Gabriel Matos}
\author[a]{Elena Busch}
\author[a]{Ki Ryeong Park}
\author[b]{Julia Gonski}
\affiliation[a]{Columbia University, 538 West 120th St, New York, NY 10027}
\affiliation[b]{SLAC National Accelerator Laboratory, 2575 Sand Hill Rd, Menlo Park, CA 94025}

\emailAdd{gabriel.pinheiro.matos@cern.ch}

\abstract{
The development of analysis methods to distinguish potential beyond the Standard Model phenomena in a model-agnostic way can significantly enhance the discovery reach in collider experiments. 
However, the typical machine learning (ML) algorithms employed for this task require fixed length and ordered inputs that break the natural permutation invariance in collision events. 
To address this, a semi-supervised anomaly detection tool is presented that takes a variable number of particle-level inputs and leverages a signal model to encode this information into a permutation invariant, event-level representation via supervised training with a Particle Flow Network (PFN). 
Data events are then encoded into this representation and given as input to an autoencoder for unsupervised ANomaly deTEction on particLe flOw latent sPacE (ANTELOPE), classifying anomalous events based on a low-level and permutation invariant input modeling. 
Performance of the ANTELOPE architecture is evaluated on simulated samples of hadronic processes in a high energy collider experiment, showing good capability to distinguish disparate models of new physics. 
}

\begin{document}
\maketitle
\flushbottom

\section{Introduction}
\label{sec:intro}

The high energy collider physics community has designated the exploration of the unknown and pursuit of new fundamental particles as key priorities for the field~\cite{p5_2014, p5_2023}. 
In parallel, the rapid development and broad adoption of machine learning (ML) for data processing and analysis has dramatically improved discovery prospects in collider physics. 
In particular, ML can be used to construct anomaly detection (AD) tools, which can classify potential hints of new physics at colliders using their incompatibility with a learned background model.
This strategy broadens the reach of existing search programs, which are primarily focused on specific extensions of the Standard Model. 
AD has been developed for a variety of applications to particle physics~\cite{BELIS2024100091}, with collider physics examples on the detection of anomalous jets~\cite{Kasieczka_2021}, dark matter~\cite{Aarrestad_2022}, event-level classification~\cite{craig2024exploringoptimaltransporteventlevel, Chekanov_2021}, or at trigger-level~\cite{Govorkova_2022, dillon}.

The performance of AD methods depends on a variety of factors.
One crucial factor is the level of supervision in training.
Most traditional ML tools rely on fully supervised training, in that they are trained over classes of data with labels to indicate the true source of an individual event as signal or background. 
This requires the generation of signal models for training samples, and generates a dependence of the tool's output classifier score on the features of the specific signals used.

Such model dependence can be mitigated through the use of less-than-supervised training methods. 
Unsupervised learning does not require any labeling of the inputs, allowing tools to train directly over unlabeled data. 
This is the type of training used with autoencoders and their variants, which have been broadly adopted for collider physics AD applications such as classification~\cite{Heimel_2019, Cerri_2019, Canelli_2022, PhysRevD.107.016002} or fast simulation~\cite{Touranakou_2022} due to their generally high performance and ease of development. 
While unsupervised learning offers the greatest generality, it can also be unstable or give results that are challenging to interpret. 
Weakly supervised training is an alternative strategy where a signal model is used in training, but with noisy or spurious labeling, allowing for training with samples that are a mixture of signal and background~\cite{Metodiev_2017}.
Another alternative is semi-supervised training, which uses correct labels but only on some of the inputs, providing another midpoint between supervised and unsupervised training that can be used for AD objectives~\cite{Park:2020pak}.
This variety of labeling options can be used in tool development, allowing a trade-off between depth of sensitivity to a narrow class of signal models, and broader but lesser sensitivity to a wide variety of models. 

A second important factor in AD tool design is the choice of input modeling, which ideally can accurately incorporate relevant features of the data, and make use of data at an early stage in collider event reconstruction to capture low-level correlations of detector signatures.
Examples of input modelings that have been used in AD for collider physics include images~\cite{Farina_2020} or sequences~\cite{Kahn_2021}.
Inherently a collider dataset is an unordered and variable-length set of particles, motivating an input modeling that is invariant to permutations of input particle ordering and can accommodate any number of inputs without zero-padding. 
Permutation invariant input modeling for high energy physics has been studied extensively, for example in the context of graph neural networks and transformers for classification~\cite{Thais:2023hmb, Abdughani_2019, Mikuni_2020, Qu_2020, Gong_2022, qu2024particletransformerjettagging, Shmakov_2022, bogatskiy2022pelicanpermutationequivariantlorentz} or construction of tracks from inner detector hits~\cite{Caillou:2815578, Burleson:2882507, Murnane:2023kfm, liu2023hierarchicalgraphneuralnetworks, choma2020trackseedinglabellingembeddedspace}.
The use of lower-level inputs also leads to higher-dimensional models~\cite{gandrakota2024robustanomalydetectionparticle, Gonski_2022}, which can better learn complex datasets but introduce a risk of overfitting or instability that must be accommodated.

The goal of this work is to bring the benefits of partial supervision in training, permutation-invariant input modeling, and low-level detector inputs to the task of AD for high energy physics, in an architecture that can be less resource-intensive or data-hungry compared to the graph neural network or transformer approach. 
This is achieved through the use of a supervised classification task to develop an intelligent embedding for data events to be used in a subsequent unsupervised training, providing customizability of the typical data-driven approach in order to have more control over what the AD model learns. 
Relatedly, other works have studied high-dimensional and permutation-invariant anomaly detection using diffusion models~\cite{Mikuni_2024} and Lorentz-equivariant autoencoders~\cite{Hao_2023}.

We introduce \textbf{ANomaly deTEction on particLe flOw latent sPacE (ANTELOPE)} to perform particle-level, variable-length, permutation-invariant AD using semi-supervised training. 
The ANTELOPE network has two steps, the first of which uses supervised training to encode inputs into a permutation-invariant basis using a Particle Flow Network (PFN)~\cite{Komiske_2019}, followed by a variational autoencoder (VAE)~\cite{kingma2022autoencodingvariationalbayes} that performs unsupervised anomaly detection over data encoded into this basis. 
This approach leads to a tool that can leverage permutation invariance and a signal model to generate a robust input modeling that results in a more performant autoencoder. 
In the following sections, the ANTELOPE architecture, training, and performance on the AD task of generic classification of beyond the Standard Model (BSM) signals are presented. 

\section{ANTELOPE Model}
\label{sec:model}

The foundation of the ANTELOPE concept for AD  arises from the encoding procedure implemented in the PFN architecture. 
The PFN is based on the Deep Sets framework for point clouds~\cite{zaheer2018deepsets}.
It consists of two stages and is trained in a supervised manner.
The first stage is an encoding network, which models an event as a list of particles and leverages particle-level data such as position and momentum to learn a per-particle representation $\Phi$.
The inputs to the PFN are given with fixed length, 
and all reconstructed final state particles are passed through the first stage of the network with a masking layer 
enforcing a length-agnostic architecture. 
A fixed-length basis $\mathcal{O}$ is then created via summation over $\Phi$ for all particles, which enforces the permutation invariance of $\mathcal{O}$. 
The encoding stage is followed by a classification network $F$ which takes signal and background events, transformed into the $\mathcal{O}$ basis, and learns binary classification using a categorical cross-entropy loss.
The encoding network, once trained, can be used independently to transform any variable-length set of low-level particle information into a fixed-length permutation invariant basis.

The ANTELOPE model then utilizes a data sample embedded via the PFN encoder stage into the fixed-length basis $\mathcal{O}$ as the input to a VAE, an extra training step which gives the classifier AD capability.
A VAE is a fixed-length architecture, which makes it challenging to apply VAEs to low-level particle data which is inherently variable in length.
By adopting the encoding stage of a pre-trained PFN, ANTELOPE is able to combine the benefits of the unsupervised autoencoder approach with low-level and permutation invariant inputs.
The use of particle inputs in the ANTELOPE architecture gives it broad applicability, as it can be used equally to model particles within jets or for a full event-level characterization. 
Developing the ANTELOPE tool thus requires two training steps: supervised training of a PFN using simulated signal and background event samples, and data-driven unsupervised training of a VAE.

The specific ANTELOPE network used in this work models events with particles described by their three-vectors of transverse momentum \pt, pseudorapidity $\eta$, and azimuthal angle $\phi$, using up to 160 of the highest \pt~particles per event. 
Events with fewer than 160 particles are zero-padded to match the required input length. 
Each feature is normalized via a Min-Max scaling to map the inputs between zero and one, preventing feature bias and aiding in model convergence.
Step 1 of the ANTELOPE training is training the PFN network on a signal-background classification task.
The first stage of the PFN network has an input dimensionality of 3 for the particle three-vector, followed by two dense fully connected layers of dimension 75, and an output $\Phi$ dimension of 64 (generating one $\Phi$ representation for each particle). 
The $\Phi$ representation is then summed over all 160 particles, resulting in one $\mathcal{O}$ 64-vector which represents information from all 160 independent particles. 
The second classification stage takes the $\mathcal{O}$ as input and thus has an input dimension of 64, followed by 3 fully connected dense layers of dimension 75, and a 2-dimensional binary classication output stage. 

Step 2 of ANTELOPE training, the VAE, takes the pre-trained PFN and utilizes its first stage to transform the same particle level information (up to 160 of the highest \pt~particles represented by their 3-vector information) of a new sample of events into the 64-vector $\mathcal{O}$ representation.
As the VAE trains without supervision, in an experimental context this step can be fully data-driven. 
This representation then becomes the input for a VAE, with a hidden layer of dimension 32, and a latent layer dimension of 12. 
The VAE loss $\mathcal{L}$ is the standard sum of the mean-squared error (MSE) between the reconstructed output and truth input, and the Kullback-Leibler divergence ($D_{\text{KL}}$) in the latent space:
\begin{equation*}
\mathcal{L} = \frac{1}{N} \sum_{i=1}^{N} | \mathcal{O}_i - \mathcal{O}^{\prime}_i |^2 + \lambda D_{\text{KL}}
\end{equation*}

\begin{figure}[!htbp]
\centering
   \includegraphics[width=0.8\textwidth]{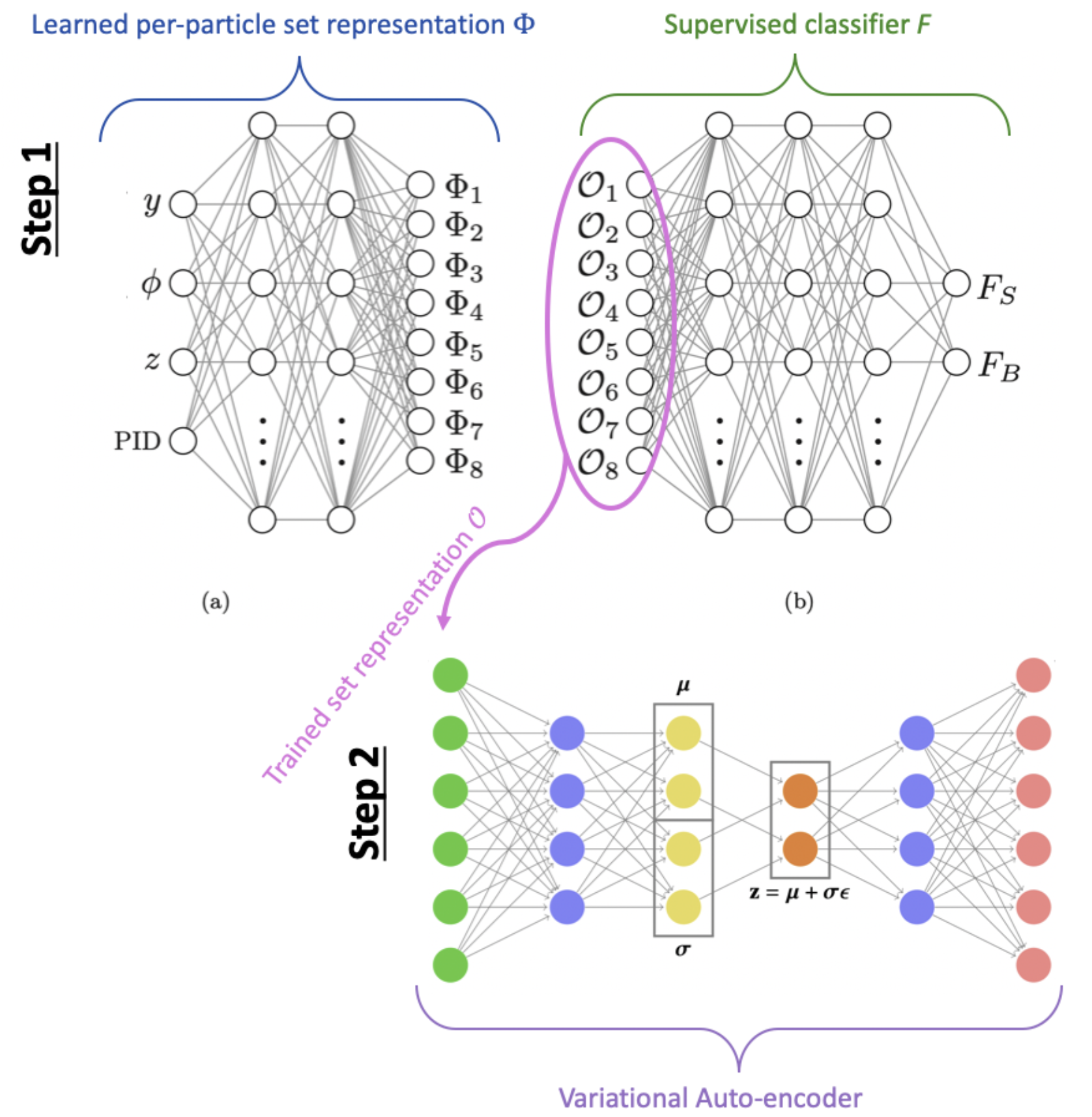}
    \caption{Diagram of the ANTELOPE architecture. The top part of the diagram shows the Particle Flow Network~\cite{Komiske_2019}, used to develop the optimal permutation-invariant basis $\mathcal{O}$ for a supervised classification task. The bottom part of the diagram shows a variational autoencoder used for anomaly detection on the Particle Flow latent space. In the case of the ANTELOPE network, the example per-particle inputs $y$, $\phi$, $z$, and PID are replaced with \pt, $\eta$, and $\phi$, which are provided for all particles in the event.  The network diagrams are for illustrative purposes only and do not reflect the dimensionality of the model used for this work, for which details are given in the text.
    \label{fig:antelope}}
\end{figure}

A diagram of the ANTELOPE architecture and its two-step training process can be seen in Figure~\ref{fig:antelope}.
The dimension of the networks were chosen via an optimization procedure, and informed by the work presented in Ref.~\cite{Komiske_2019}. 
The performance was observed to be robust across many small variations of the architecture.

\section{Samples \& Training}
\label{sec:training}

To assess the ability of ANTELOPE to classify BSM signals from background in a model-independent way, the network is trained and evaluated over samples of simulated collider events with hadronic final states. 
The samples used are from the LHC Olympics (LHCO)~\cite{Kasieczka_2021}, a community competition where different AD tools were applied to determine the nature of an unknown signal in ``Black Box" datasets. 
The datasets provided include a background of Standard Model multijet processes, two known R\&D BSM signals to be used for tool development, and three Black Boxes. 
Since the competition, the Black Boxes have been unblinded and the underlying processes were revealed, enabling their use for ANTELOPE performance evaluation. 

\begin{figure}[tbh]
\centering
\includegraphics[width=0.98\textwidth]{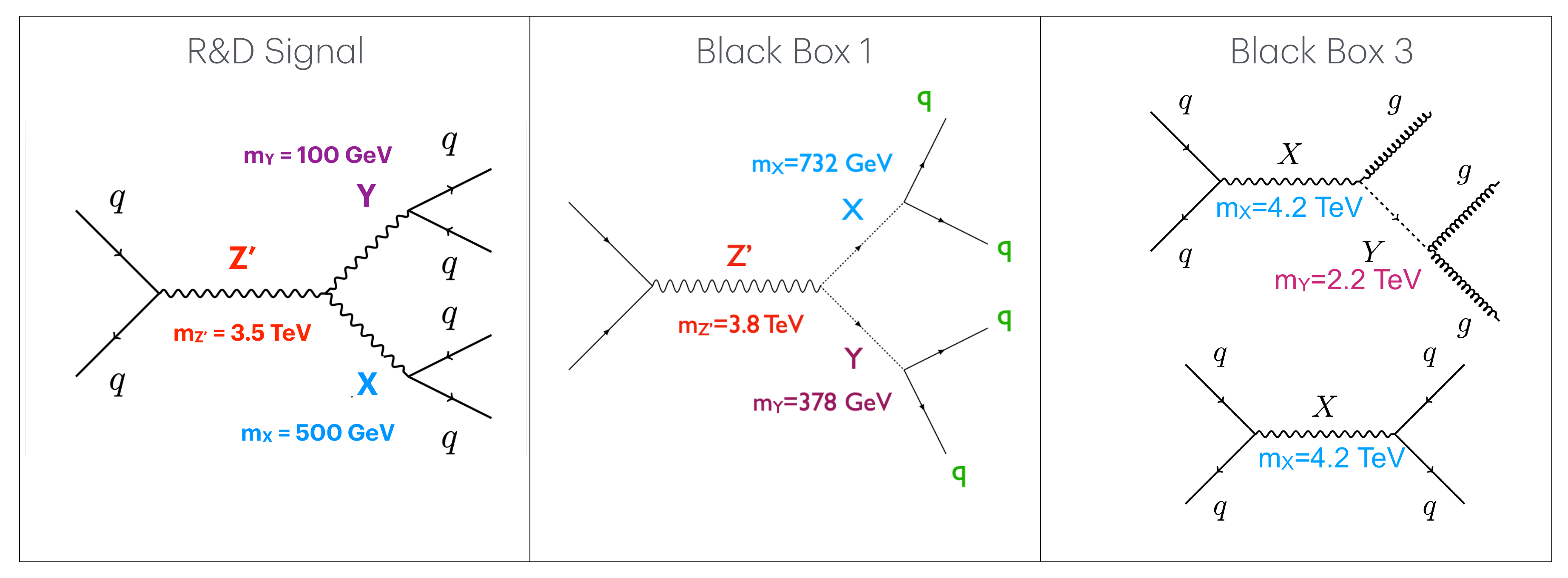}
\caption{\label{fig:feynmanDiag} Feynman diagram of signal processes used in the training and evaluation of the ANTELOPE network, from Ref.~\cite{Kasieczka_2021}. Signals include a heavy resonance decaying to two two-prong jets (left), another two-prong signal with different particle masses (middle), and a signal with a heavy resonance and two decay modes (right). The three-prong R\&D signal is the same as the two-prong, but with the $X$ and $Y$ particles each decaying to 3 quarks.}
\end{figure}

The R\&D signals consist of a simulated process of $s$-channel heavy \zp~production, where the \zp~
decays to daughters $X$ and $Y$ which subsequently decay to quarks. 
The heavy mass of the \zp~leads to boosted $X$ and $Y$ decays, such that they are best reconstructed as a single large-radius jet with substructure. 
Two of these signals are provided for algorithm development, one where the $X$ and $Y$ each decay to two quarks leading to two-prong substructure, and another with three quark decays leading to three-prong substructure. 
Black Box 1 contains a signal that is similar to that of the R\&D samples, with a \zp~to $X$ and $Y$ process with two quark boson decays, but with different BSM particle masses than the R\&D signals. 
Black Box 2 contains no signal, only Standard Model multijet background processes.
Black Box 3 contains a signal of a heavy resonance with two decay modes, one to two quarks and another to a gluon and another heavy particle $Y$, both of which must be detected for sensitivity.
Figure~\ref{fig:feynmanDiag} provides Feynman diagrams of the key signal processes. 

In total, six samples are used in this analysis: the background, two R\&D signals, and three Black Boxes (one of which contains only background processes). 
To verify AD capability, the ANTELOPE tool must be trained using one of the R\&D signal models, but results in a classifier that can also identify the Black Box signals as anomalous without advance knowledge of their distinguishing features.

\begin{figure}[tbh]
\centering
\includegraphics[width=0.45\textwidth]{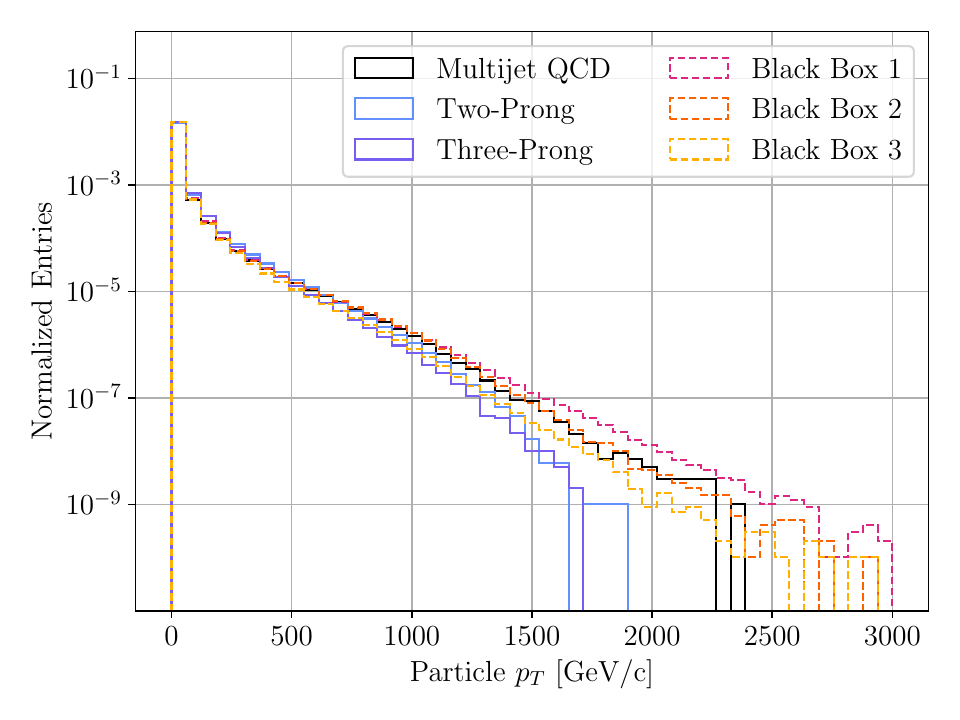}
\includegraphics[width=0.45\textwidth]{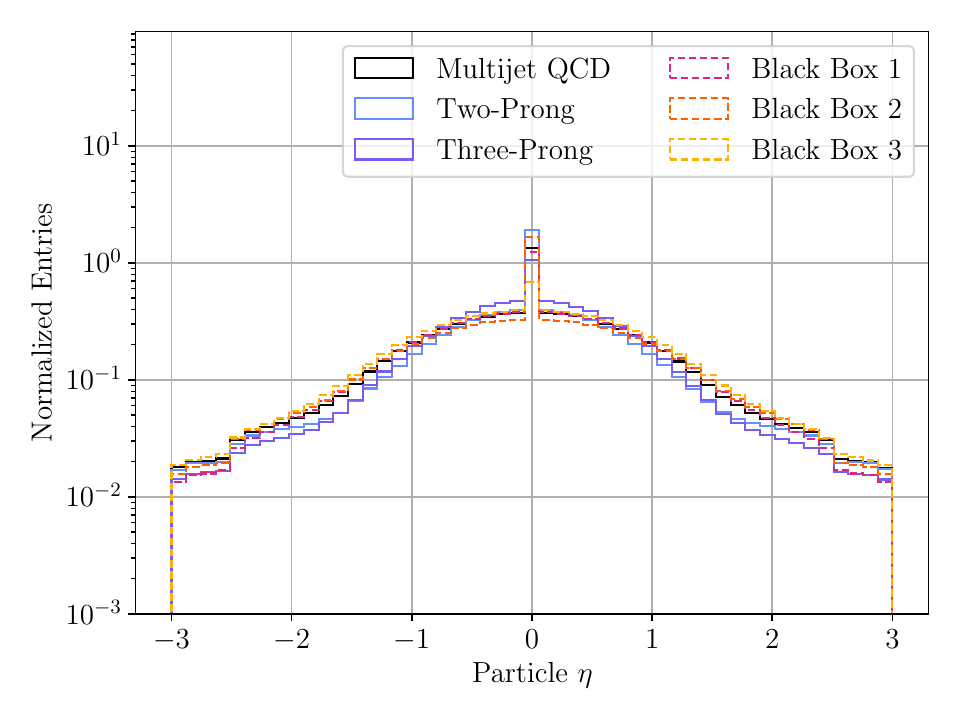}
\includegraphics[width=0.45\textwidth]{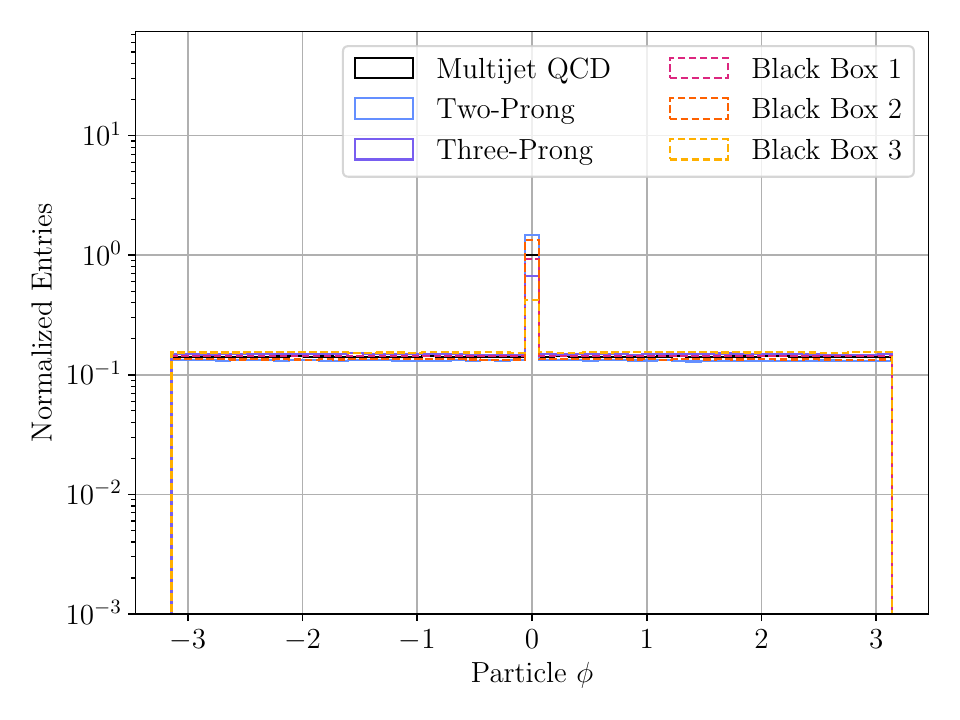}
\caption{\label{fig:inputs} Distributions for the six analysis samples of the input variables used in PFN and ANTELOPE training, namely the \pt~(left), $\eta$ (center), and $\phi$ (right) for the 160 leading particles in each event used in training including all analysis sample events. The excess of events at values of 0 indicate the use of zero-padding if the event has fewer than 160 particles.}
\end{figure}

\begin{figure}[tbh]
\centering
\includegraphics[width=0.45\textwidth]{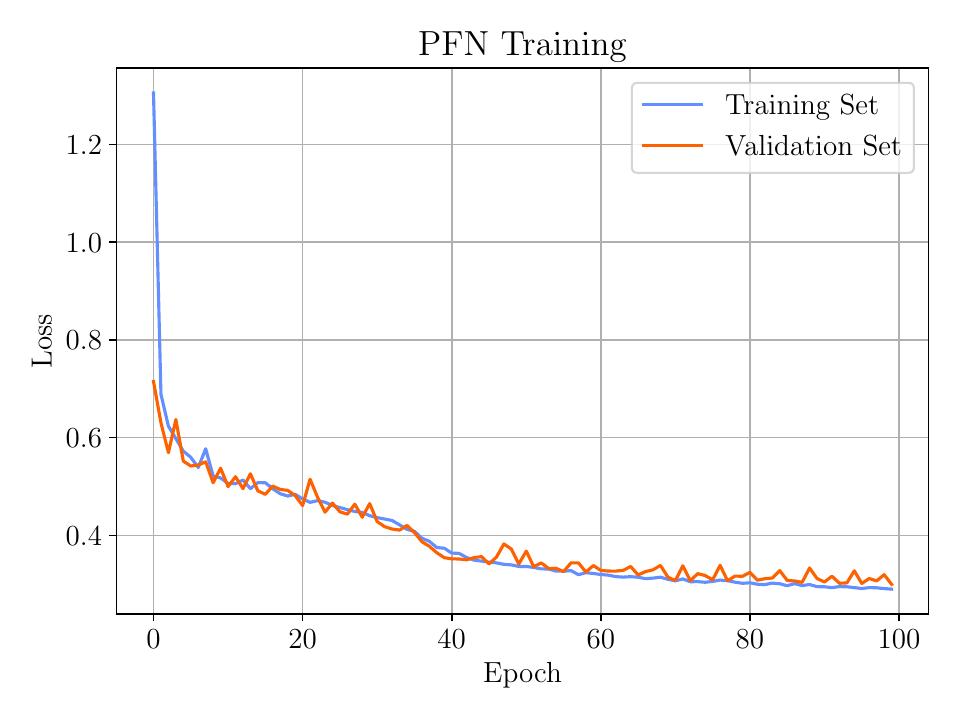}
\includegraphics[width=0.45\textwidth]{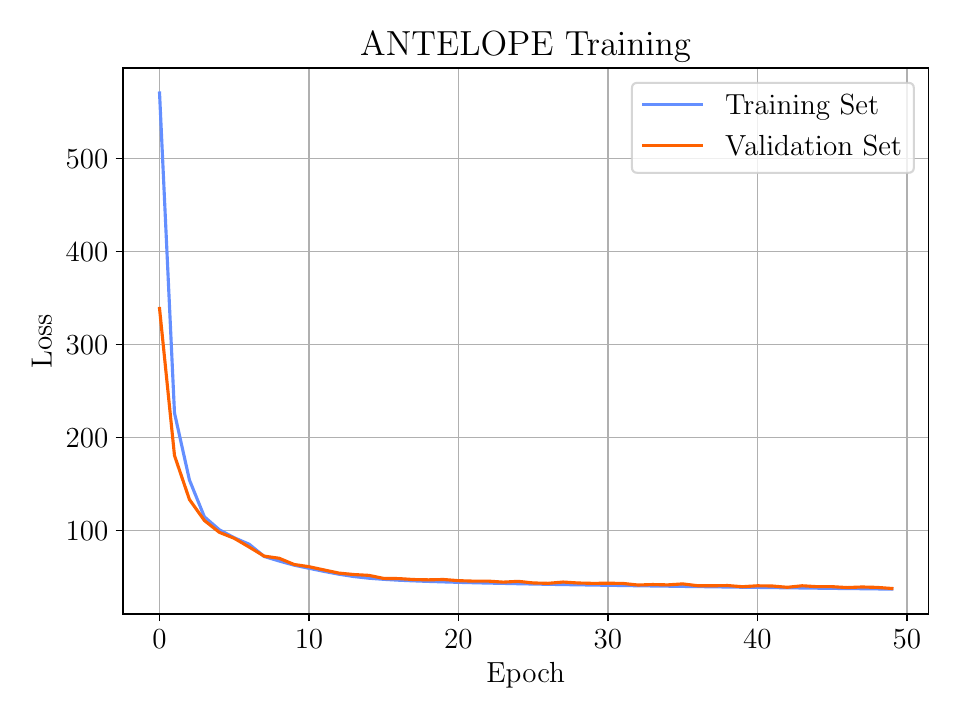}
\caption{\label{fig:loss} Distributions of the loss during training for the Step 1 PFN model (left) and Step 2 ANTELOPE VAE model (right), including both training and validation samples.}
\end{figure}

Each event in the LHCO samples is given as a list of up to 700 particles, which are described by their \pt, $\eta$, and $\phi$. 
Distributions of these variables for the six analysis samples is given in Figure~\ref{fig:inputs}, where only the 160 highest $p_T$ particles used in the training are included.
Background and R\&D signal analysis samples are split into orthogonal training and test datasets comprising 80\% and 20\% of the sample, respectively.
For the other analysis samples not involved in the training, the test set comprises the full set of generated events: 100,000 events for the three-prong R\&D sample, 834 signal events for Black Box 1, 1 million events for Black Box 2, and 1,200 dijet and 2,000 trijet signal events for Black Box 3. 
The PFN model used to generate the embedding for the VAE input is trained for 100 epochs using 80,000 events of background and the two-prong R\&D signal each.
The VAE is trained over 80,000 orthogonal events from the same background sample, meant in this case to represent data, encoded into the pre-trained PFN $\mathcal{O}$ basis for 50 epochs. 
Figure~\ref{fig:loss} shows the output PFN and ANTELOPE loss as a function of epoch during training in both training and validation samples, showing good stability. 

\section{Results}
\label{sec:results}

The performance of ANTELOPE is assessed through its ability to discriminate the various LHCO signals from the multijet background processes.
In order to verify that the ANTELOPE tool is better suited for AD than a supervised PFN, we evaluated both ANTELOPE and the PFN over the test sets of the six analysis samples and compared their performance. 
ANTELOPE should be able to better classify the varying signal models from the background than what is achievable with the PFN trained over a single signal model.
The output loss of ANTELOPE is used as an anomaly score that can be thresholded to define a signal-enriched region for the evaluation of model sensitivity. Similarly, the PFN's output loss is used as a control score to determine the extent of model-independence achieved with ANTELOPE.
Black Box 2, containing only background events, is also evaluated to ensure the network does not memorize the training sample and can provide good anti-background tagging capability on alternate background samples. 

The performance of the PFN and ANTELOPE models is quantified by the receiver operating characteristic (ROC) area-under-curve (AUC), evaluated for each of the analysis samples.
The significance improvement characteristic (SIC) is also considered, defined as the ratio of the true positive rate (TPR) to the square root of the false positive rate (FPR). This gives a proxy for the factor by which signal significance is improved for a given threshold on the output model loss. 
These metrics are used to compare the fully supervised PFN approach to the AD capability provided by ANTELOPE. 

\begin{figure}[tbh]
\centering
\includegraphics[width=0.45\textwidth]{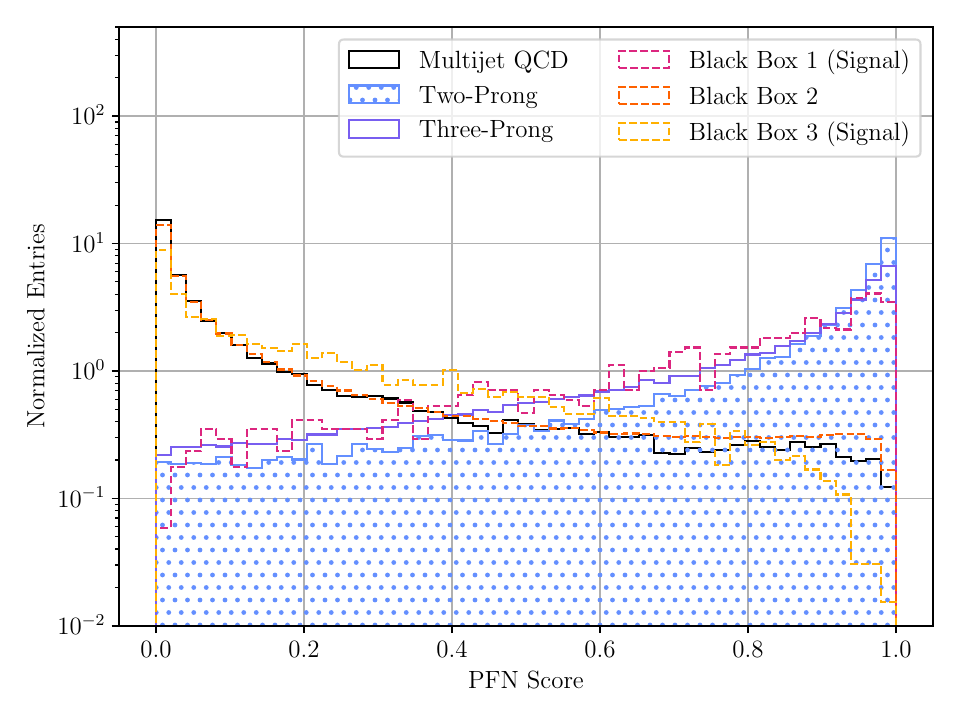}
\includegraphics[width=0.45\textwidth]{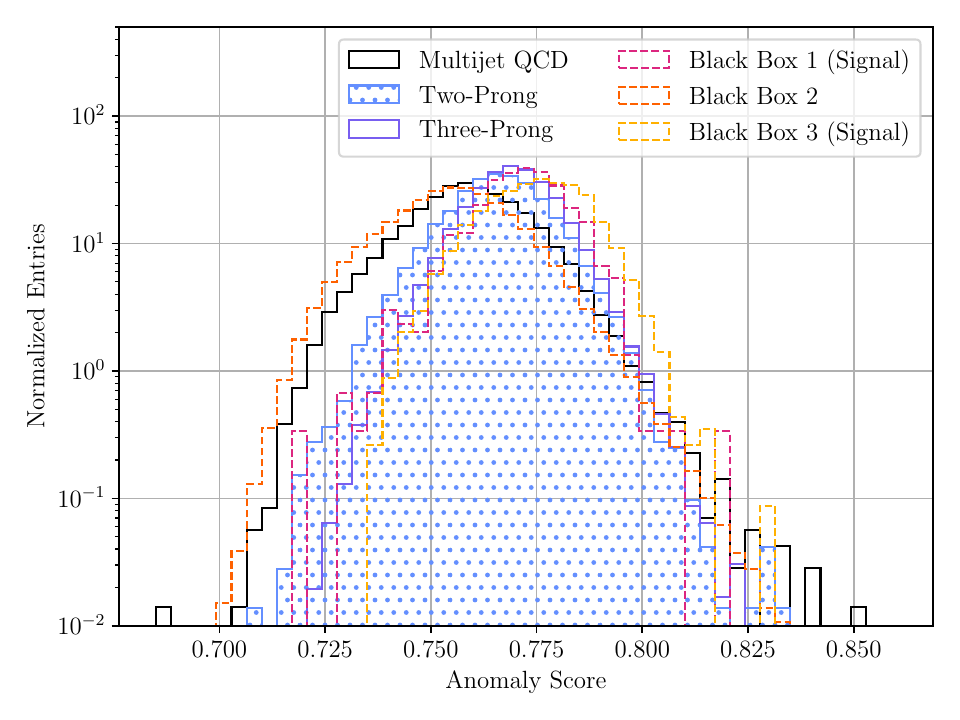}
\caption{\label{fig:scores} The classifier score, determined by the output loss of the PFN (left) and ANTELOPE (right) models when evaluated over test sets of the six analysis samples. Higher values indicate more signal-like characterization. A dot pattern indicates the signal sample that was used to train the PFN.}
\end{figure}

Figure~\ref{fig:scores} compares the output loss distributions for both the PFN and ANTELOPE models across the test sets of the six analysis samples.  
The benefit of a high-dimensional input modeling is such that while there is no strong discrimination between signal and background among these low-level distributions shown in Figure~\ref{fig:inputs}, the correlations between them considering $\mathcal{O}$(100) particles in each event provides good classification power. 
This is demonstrated by the good discrimination achieved by the PFN when evaluated over the two-prong R\&D signal (the same signal used in the supervised training).
It also generalizes well to the three-prong and Black Box 1 signal, which share the two-prong signature of jet substructure, giving some indication of what the PFN is able to learn to distinguish signals from background. 
In comparison, the output ANTELOPE score distribution has a poorer reconstruction quality for all four signal samples, pushing them into the tail of the loss distribution, while maintaining a lower score for the two background samples, enabling some shape difference that can be leveraged for an AD classifier.

\begin{figure}[tbh]
\centering
\includegraphics[width=0.45\textwidth]{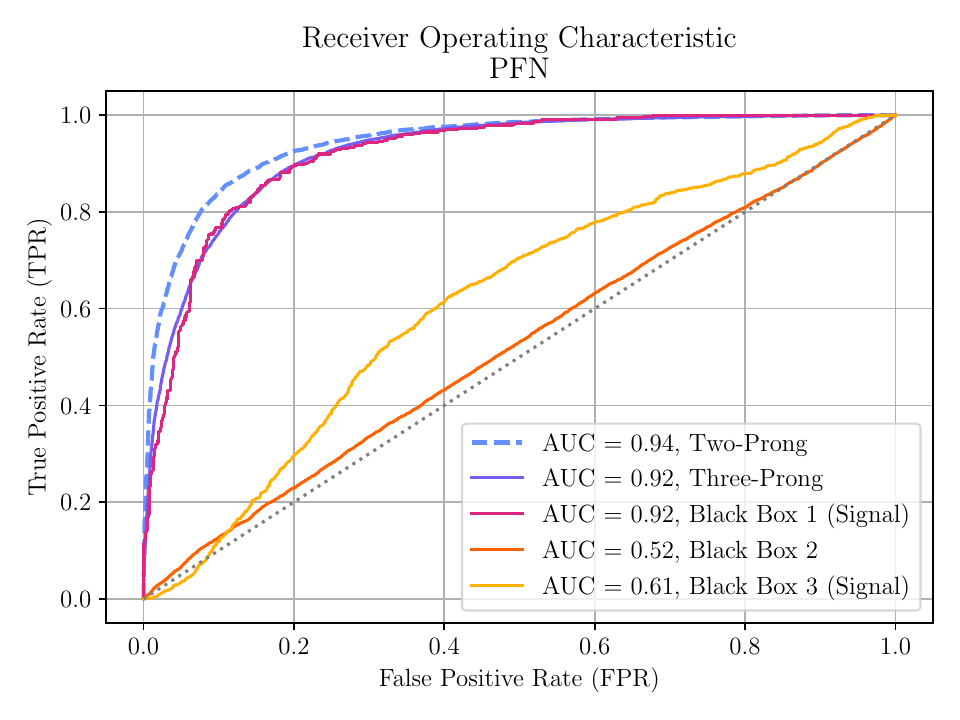}
\includegraphics[width=0.45\textwidth]{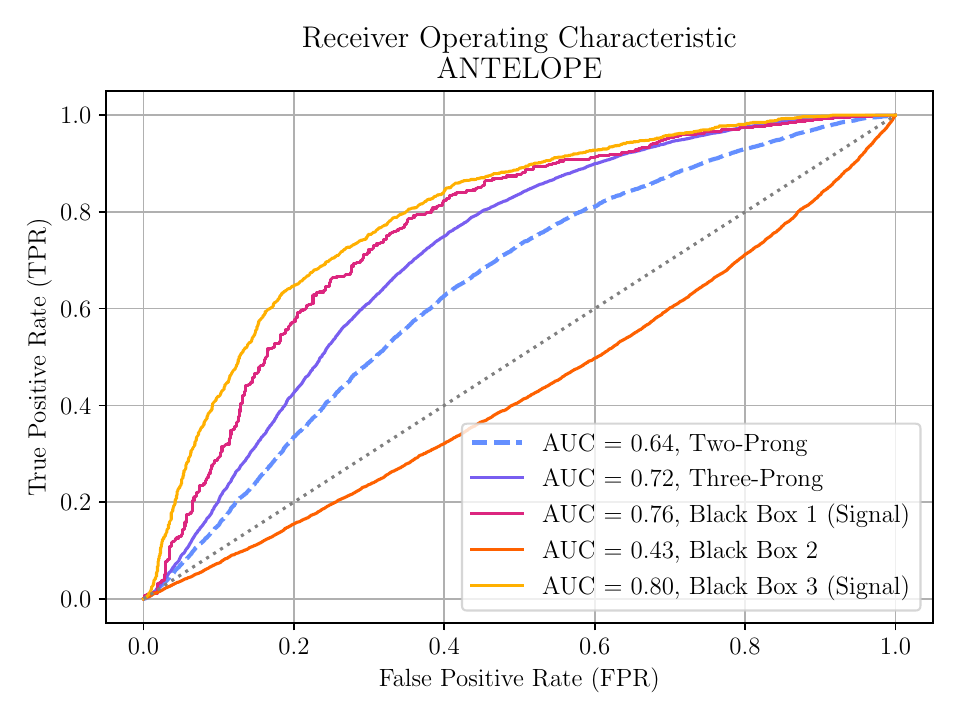}
\includegraphics[width=0.45\textwidth]{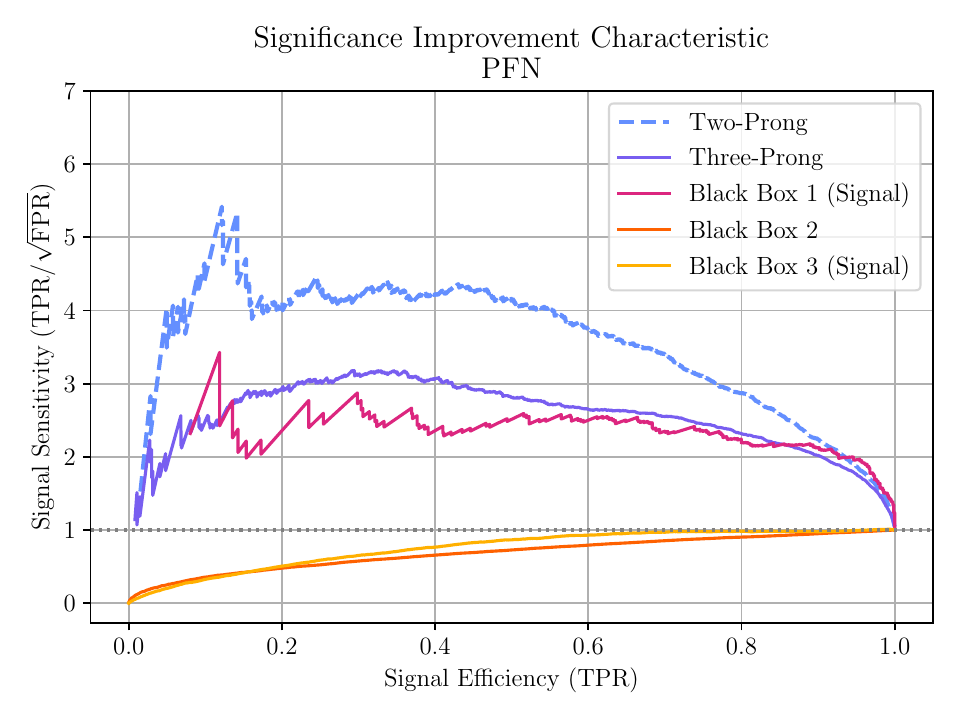}
\includegraphics[width=0.45\textwidth]{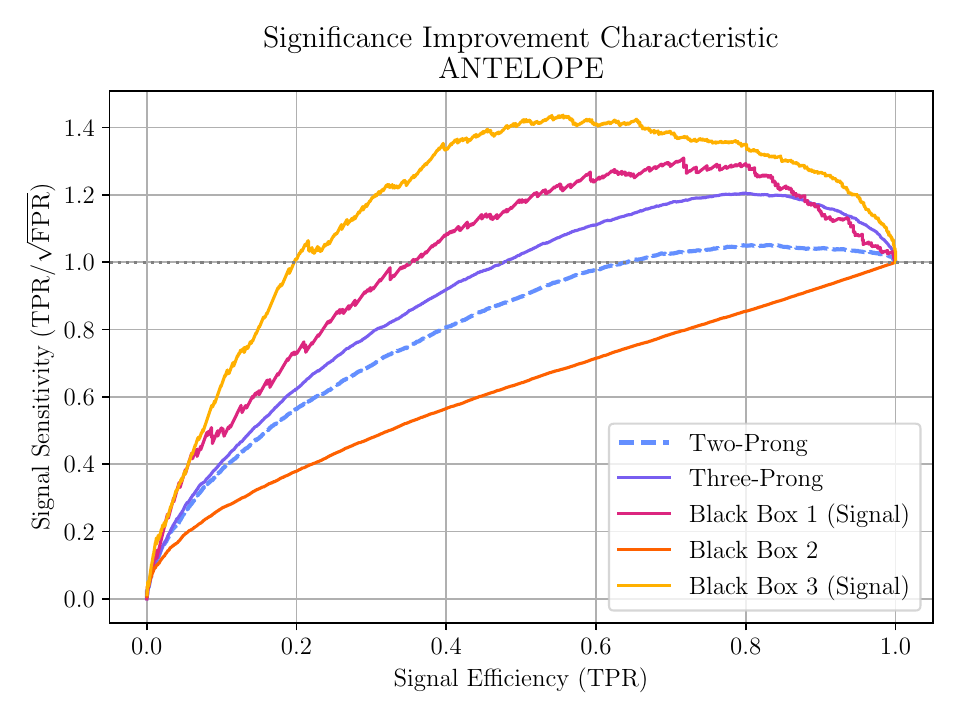}
\caption{\label{fig:sic} The receiver operating characteristic (ROC, top) and the significance improvement characteristic (SIC, bottom), for the PFN (left) and ANTELOPE (right) networks. The SIC is defined as TPR divided by $\sqrt{\text{FPR}}$. A black dashed line indicates the performance of a random guess classifier. The addition of the unsupervised training step in the ANTELOPE architecture enhances the generality of the model loss in acting as a model-independent classifier.}
\end{figure}

Figure~\ref{fig:sic} compares the ROC and SIC curves for both the PFN and ANTELOPE models across the six analysis sample test sets. 
The value on the $y$-axis can be considered as the factor by which the signal sensitivity would be improved by thresholding on the anomaly score at the corresponding $x$-axis value. 
As is reflected in Figure~\ref{fig:scores}, using only the supervised PFN as a classifier gives the greatest sensitivity enhancement on the two-prong model used in training and the similar substructure models in the three-prong and Black Box 1 samples.
However, it is unable to provide any sensitivity enhancement for Black Box 3, with multiple resonant decay modes. 

In comparison, ANTELOPE has lower sensitivity to each individual signal, except for Black Box 3, which it ranks as the most anomalous of all samples. 
This supports the hypothesis that the addition of the VAE training generates AD capability, as the ANTELOPE network can identify an unusual model not classified as signal-like by the supervised approach. 
In this way, it is observed that ANTELOPE provides sensitivity to a broad set of signal models with a single score, albeit with a loss in performance compared to the supervised approach for the narrow class of resonant particles with substructure.
In the LHC Olympics competition, the Black Box 3 signal was not detected by any method with its true mass and competing decay modes~\cite{Kasieczka_2021}. 
This indicates the benefit of a supervision-derived input embedding to enhance traditional autoencoder-based approaches.
Neither the PFN nor the ANTELOPE considers Black Box 2 as signal-like or anomalous, indicating robust networks that are resistant to overtraining. 

To test the sensitivity of the learned PFN basis $\mathcal{O}$ to a particular signal model, an alternate ANTELOPE network was developed that trained on the three-prong R\&D signal instead of the two-prong. A similar trend is observed as for the two-prong training shown above, where the PFN tool alone provides the highest performance for the signal used in training and related models, but the addition of the unsupervised component in the ANTELOPE allows sensitivity to the Black Box 3 signal.
This indicates that the observed benefits of ANTELOPE are not highly dependent on the signal model used to develop the PFN embedding, enhancing its robustness in BSM search applications.

The semi-supervised nature of the ANTELOPE method allows it to take advantage of simulation without depending on it directly in collider analyses. 
While the supervised training of the PFN uses simulated signal and background models, the VAE stage trains in an unsupervised way and can thus be trained directly over data. 
The strength and generality of this approach are evidenced by the ANTELOPE network's ability to classify the challenging Black Box 3 signal as anomalous, and with the best performance of any signal models tested.
Further, the low-level input modeling broadens the ANTELOPE usage to classification tasks in a variety of contexts such as jet-level, event-level, or for non-resonant physics.
While the signals presented here represent a relatively specific class of BSM physics, with resonances and hadronic final states, future efforts can consider different detectors or more challenging low-level objects such as displaced or disappearing tracks arising from long-lived BSM particles.

\section{Conclusions}
\label{sec:conclusions}

The ANTELOPE method is presented as a way to leverage the benefits of less-than-supervised ML training and intelligent permutation-invariant input embeddings for the task of classifying anomalous BSM signatures in collider datasets. 
The method comprises a pre-training stage where a PFN is developed to classify background from a specific signal model, and the permutation-invariant basis developed through this pre-training is then used to embed background-only events for the unsupervised signal model-free training of a VAE.
The ANTELOPE network is found to give broader sensitivity across several signal models with varying distinguishing characteristics as compared to the PFN alone, which outperforms the semi-supervised approach when evaluating over a single class of signals but fails to generalize to more challenging models. 
To fully understand the trade-off between model performance and size/complexity, further studies could compare ANTELOPE to transformer or graph approaches with a potential target for online implementation. 
This and other future studies with low-level and ML-derived input modelings can further assist classical autoencoder approaches for robust AD programs at colliders with unique and exciting discovery potential.

\appendix
\section*{Code Availability}

The code for the ANTELOPE model is publicly available at \\ \url{https://github.com/gabrielpmatos/ANTELOPE}.

\acknowledgments

The authors acknowledge Mike Tuts and Daniel Murnane for discussions and input. 
GM, EB, and KP are supported by the National Science Foundation under Grant No. PHY-2310080. 
JG is supported by the U.S. Department of Energy under contract number DE-AC02-76SF00515.


\bibliographystyle{JHEP}
\bibliography{biblio.bib}


\end{document}